\documentclass[aps,prl,twocolumn,reprint,superscriptaddress,citeautoscript,floatfix,longbibliography]{revtex4-1}

\usepackage[dvipdfmx]{graphicx}
\usepackage{amsmath,amssymb}
\usepackage{fixmath}
\usepackage[normalem]{ulem}
\usepackage{bm}
\usepackage{physics}

\usepackage{color}

\usepackage{comment}

\begin{document}

\title{Gyromagnetic bifurcation in a levitated ferromagnetic particle}

\author{T. Sato}
\author{T. Kato}
\affiliation{Institute for Solid State Physics, University of Tokyo, Kashiwa 277-8581, Japan}
\author{Daigo Oue}
\affiliation{%
Kavli Institute for Theoretical Sciences, University of Chinese Academy of Sciences, Beijing, 100190, China.
}%
\affiliation{%
The Blackett Laboratory, Imperial College London, London SW7 2AZ, United Kingdom
}%
\affiliation{%
Instituto Superior T\'{e}cnico, University of Lisbon, 1049-001 Lisboa, Portugal
}%

\author{M. Matsuo}
\affiliation{%
Kavli Institute for Theoretical Sciences, University of Chinese Academy of Sciences, Beijing, 100190, China.
}%
\affiliation{CAS Center for Excellence in Topological Quantum Computation, University of Chinese Academy of Sciences, Beijing 100190, China}
\affiliation{Advanced Science Research Center, Japan Atomic Energy Agency, Tokai, 319-1195, Japan}
\affiliation{RIKEN Center for Emergent Matter Science (CEMS), Wako, Saitama 351-0198, Japan}

\date{\today}

\begin{abstract} % 600 characters including spaces
We examine mechanical rotation of a levitated magnetic particle that is induced by ferromagnetic resonance under microwave irradiation. We show that two stable solutions appear in a certain range of parameters by bifurcation when the rotation frequency is comparable to the microwave frequency. This phenomenon originates from the coexistence of the Barnett and the Einstein-de Haas effects. We also reveal that this measurement is sensitive to the strength of the spin-rotation coupling. Our work provides a platform for accessing a microscopic relaxation process from spin to macroscopic rotation.
\end{abstract}

\maketitle

{\it Introduction.---} The discovery of gyromagnetism~\cite{Barnett1915,Einstein_deHaas_1915}, i.e., the interconversion between spin and mechanical rotation, was a milestone in magnetism research, because it revealed the origin of magnetism to be the intrinsic angular momentum of electrons, which classical physics cannot explain, even before the establishment of quantum mechanics. Originally, gyromagnetic effects were investigated in bulk magnetic materials with aim of determining their gyromagnetic ratios~\cite{Scott1962}. Recently, gyromagnetic effects have been recognized as universal phenomena, as demonstrated in various systems at scales ranging from those of condensed matter~\cite{harii2019spin,Mori2020,Matsuo-PRL-2011,matsuo2013mechanical,hirohata2018magneto,Takahashi2016,Takahashi2020,Kazerooni2020,Kazerooni2021,Kobayashi2017,kurimune2020highly,tateno2020electrical,tateno2021einstein} to those of particle physics~\cite{Adamczyk2017,adam2018global,adam2019polarization,acharya2020evidence,adam2021global}. They have also provided powerful tools with which to measure and control both the mechanical and magnetic degrees of freedom~\cite{Wallis2006,Zolfagharkhani-NatNano-2008,imai2018observation,imai2019angular,harii2019spin,Mori2020,Dornes2019,izumida2022einstein}.

Indeed, the Barnett effect (conversion of angular momentum of mechanical rotation into spin~\cite{Barnett1915}) has been utilized to identify the angular momentum compensation temperature of ferrimagnets~\cite{imai2018observation,imai2019angular} and to generate spin current from rigid-body rotation~\cite{Matsuo-PRL-2011,matsuo2013mechanical,hirohata2018magneto}, surface acoustic waves~\cite{Kobayashi2017,kurimune2020highly,tateno2020electrical}, and vorticity in fluids~\cite{Takahashi2016,Takahashi2020,Kazerooni2020,Kazerooni2021}. On the other hand, the Einstein--de Haas effect (mechanical torque generated from spin polarization~\cite{Einstein_deHaas_1915}), which is abbreviated as the EdH effect hereafter, has been utilized to measure the faint torque caused by a single electron spin flip~\cite{Zolfagharkhani-NatNano-2008}, identify the gyromagnetic ratio of a nanomagnetic thin film~\cite{Wallis2006}, and reveal the demagnetization process in ferromagnets on sub-picosecond time scales~\cite{Dornes2019}.
 
Very recently, it has been demonstrated, in a solid-state device, that the Barnett and EdH effects can coexist in the GHz-frequency regime~\cite{tateno2021einstein}, although the Barnett and EdH effects have been treated independently so far. Through these two routes, angular momenta can bounce back and forth between magnetic and mechanical degrees of freedom in a single system; hence, their coexistence may bring out rich physics and possibly highlight the microscopic features of the spin-rotation coupling, the fundamental coupling between spin and mechanical angular momentum of a rotating body~\cite{hehl1990inertial,Frohlich1993}. However, solid-state platforms involve various excitations; hence, the gyromagnetic phenomena are damped by and could even be buried in, e.g., spin and charge transport and impurity scattering. In this sense, it is crucial to construct a platform that is detached from those relaxation paths and enables high-frequency, stable rotation.

Levitated optomechanics, inspired by laser tweezers, levitation, and cooling~\cite{ashkin1970acceleration,ashkin1976optical,ashkin1986observation,chu1985three,vuletic2000laser}, can provide such a platform. Stable levitation of small particles has been demonstrated under a high vacuum through the use of, e.g., optical and radio-frequency forces~\cite{gonzalez2021levitodynamics}. In particular, with optical forces controlled by parametric feedback, sub-kelvin cooling of center-of-mass motion of small particles has been demonstrated~\cite{li2011millikelvin,gieseler2012subkelvin,jain2016direct}. By combining the feedback cooling scheme with the cavity cooling technique~\cite{chang2010cavity,millen2015cavity}, even zero-point fluctuation of the center of mass has been revealed~\cite{tebbenjohanns2020motional}. Since motion of the center of mass can be significantly suppressed, high-frequency rotation ($\sim \mathrm{GHz}$) of small particles in a high vacuum has been studied recently~\cite{Reimann2018,Ahn2018}. 
The levitation of ferromagnets, which is necessary for our setup, has been studied theoretically in previous works~\cite{cimurs2013three,cimurs2013dynamics,usadel2017dynamics,usadel2015dynamics,usov2015magnetic,usov2012dynamics,keshtgar2017magnetomechanical,lyutyy2019uniform,lyutyy2019dissipation} and has achieved also experimentally~\cite{Huillery2020,Supplement}.
Though the levitated optomechanical measurements are thus best suited for investigating the interplay between the Barnett and EdH effects,
%The levitation of ferromagnets, which is necessary for our setup, has also already achieved \cite{Huillery2020,Supplement}. Though the levitated optomechanical measurements are thus best suited for investigating the interplay between the Barnett and EdH effects,
%%in the GHz-frequency regime, 
such an application has been overlooked so far.
%\blue{The setups similar to ours are considered in many previous works \cite{cimurs2013three,cimurs2013dynamics,usadel2017dynamics,usadel2015dynamics,usov2015magnetic,usov2012dynamics,keshtgar2017magnetomechanical,lyutyy2019uniform,lyutyy2019dissipation}, but the coexistence of Barnett/EdH effects are not considered at all, and the gyromagnetic phenomena have not been investigated enough.}

\begin{figure}[tbp]
\includegraphics[width=\linewidth]{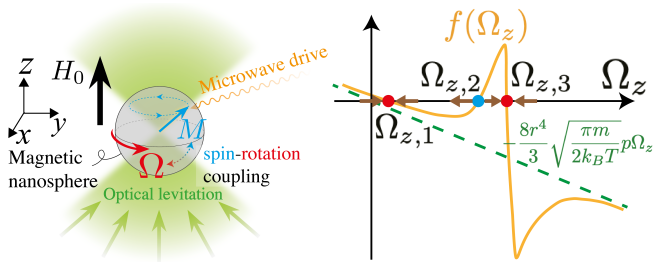}
 \caption{(a) Schematic picture of the model. The magnetic nanosphere is optically levitated. The static magnetic field $H_0$ is applied in the $z$ direction and the circularly-polarized microwave field in the $xy$ plane. They induce precession of the magnetization ${\bm M}$ and macroscopic rotation ${\bm \Omega}$ about the $z$ direction via the spin-rotation coupling. (b) Net angular momentum gain $f(\Omega_z)$. There are three steady-state solutions, $\Omega_{z,i}$ ($i=1,2,3$). For $\Omega_{z,1}$ and $\Omega_{z,3}$ (red points), an infinitesimal change in $\Omega_z$ produces a restoring torque, and thus, the solutions become stable. In contrast, an infinitesimal deviation subsequently grows for $\Omega_{z,2}$, and thus, that solution is unstable.}
 \label{fig:geometry}
\end{figure}

In this Letter, we propose a levitated optomechanical setup to directly probe angular-momentum transfer between a spin system and mechanical rotation via Gilbert damping with high precision. We consider a small levitated particle and study its uniaxial rotation under microwave irradiation in a ferromagnetic resonance (FMR) experiment (see Fig.~\ref{fig:geometry}~(a)). We calculate the steady-state rotation frequency by balancing energy injection from microwaves and energy loss from air resistance and conclude that the particle can be rotated with a high rotation frequency up to GHz order in a vacuum. We show that the steady-state solution exhibits bifurcation when the rotation frequency is comparable to the microwave frequency, which is the requirement for the strong Barnett effect, in other words, coexistence of the Barnett and the EdH effects. 
The present setup enables us to sensitively measure the $g$-factor in the spin-rotation coupling. 
Our result illustrates the usefulness of levitated optomechanical techniques in the study of gyromagnetism.

{\it Dynamics of gyromagnetic systems.---} We consider a spherical ferromagnetic levitated particle of radius $r$, in which the motion of the center of mass is suppressed. For simplicity, the particle is regarded as a rigid body with a moment of inertia, $I=2m_\mathrm{ptc} r^2/5$, where $m_\mathrm{ptc}$ is the particle mass. We assume that the magnetization of the particle, ${\bm M}$, is initially directed in the $z$-direction by an external static magnetic field. We consider a ferromagnetic resonance (FMR) experiment in which external microwaves irradiate a sample particle~\cite{Kittel1948}. In this experiment, the angular momentum of the excited spins is transferred via the Gilbert damping to the rigid-body rotation of the particle and therefore the particle starts to rotate around the $z$-axis. Its rotation frequency vector is denoted as ${\bm \Omega}=(0,0,\Omega_z)$.

For the steady state, the Hamiltonian in the rotating frame fixed to the particle is written as \cite{hehl1990inertial,Frohlich1993,Matsuo2013}
\begin{align}
{\cal H} = -(\mu_0\gamma{\bm H}+g_{\rm SR} {\bm \Omega}) \cdot \hbar {\bm S}^{\rm tot}.
%-\hbar \mu_0\gamma{\bm H}\cdot{\bm S}^{\rm tot}-\hbar g_{\rm SR} {\bm \Omega}\cdot{\bm S}^{\rm tot}.
\label{eq:Hamiltonian}
\end{align}
The first term describes the Zeeman energy, where ${\bm S}^{\rm tot}$ is the total spin of the particle, $\mu_0$ is the vacuum permeability, $\gamma$ ($<0$) is the gyromagnetic ratio, and ${\bm H}$ is the magnetic field in the rotating frame, 
\begin{align}
\label{eq:H}
{\bm H}=(h\cos(\omega-\Omega_z) t,h\sin(\omega-\Omega_z) t,H_0).
\end{align}
Here, $H_0$ is a static magnetic field and $h$ ($\ll H_0$) and $\omega$ are the amplitude and frequency of the microwaves, respectively. The second term of the Hamiltonian (\ref{eq:Hamiltonian}) describes the spin-rotation coupling that explains the Einstein-de Haas effect~\cite{Einstein_deHaas_1915} and the Barnett effect~\cite{Barnett1915}. Note that the $g$-factor for the spin-rotation coupling, $g_{\rm SR}$, generally deviates from one~\cite{Supplement,Matsuo2013}.

The magnetization of the particle is given as ${\bm M} = \hbar \gamma {\bm S}^{\rm tot}/V$, where $V=4\pi r^3/3$ is the volume of the particle. 
The Landau–Lifshitz–Gilbert (LLG) equation is given as
%From the Hamiltonian, the Landau–Lifshitz–Gilbert (LLG) equation is derived as
\begin{align}
\label{LLG}
\dot{{\bm M}} =
{\bm M}\times (\mu_0\gamma {\bm H}+g_{\rm SR}{\bm \Omega})+\frac{\alpha}{M_0}{\bm M}\times
\dot{{\bm M}},
\end{align}
where $M_0=|{\bm M}|$, $\dot{{\bm M}}={\rm d}{\bm M}/{\rm d}t$, and $\alpha$ is the Gilbert damping constant. The steady-state solution of the LLG equation can be obtained by assuming $\dot{M_z}=0$ and $(M_{x},M_y) = M(\cos\left[(\omega -\Omega_z)t+\phi\right],\sin\left[(\omega -\Omega_z)t+\phi\right])$
and using $M_0^2=M^2+M_z^2$. For this steady state, the $z$ component of the gain rate of the angular momentum from the spin system is given as
\begin{align}
\label{eq:FMR}
& \left[{\bm M}\times\mu_0\gamma {\bm H} \right]_z= -\frac{\alpha}{M_0}(\omega-\Omega_z)M^2.
\end{align}
This coincides with the Gilbert damping term $\alpha\left[{\bm M}\times\dot{\bm M}\right]_z/M_0$, indicating that the gain in angular momentum from the microwaves is lost through Gilbert damping. This spin relaxation causes continuous transfer of the angular momentum to the rigid-body angular momentum.

{\it Magnetically driven rigid-body rotation.---} The equation for the time evolution of the angular momentum in the present system is given as~\cite{Supplement}
\begin{align}
\label{eq:conservation}
\dv{}{t}\left[I\Omega_z+g_{\rm SR}\hbar S^{\rm tot}_z\right]=\Gamma_{\rm in}+\Gamma_{\rm air},
\end{align}
where $\Gamma_{\rm in}=\mu_0\gamma \hbar \left[{\bm S}^{\rm tot}\times{\bm H}\right]_z$ is the torque supplied from the microwaves and $\Gamma_{\rm air}$ is the torque due to the air resistance (discussed later). When the Gilbert damping term is included phenomenologically as in the last term of Eq.~(\ref{LLG}), a steady state, $\dot{S}_z=0$, arises from the balance between the Gilbert damping and the gain in angular momentum from the microwaves, Eq.~(\ref{eq:FMR}). Then, Eq.~(\ref{eq:conservation}) leads to the equation of motion for the steady-state rotation frequency $\Omega_z$. 

The steady-state rotation of the particle receives torque from the surrounding air. In this paper, we focus on the molecular flow region in which $\lambda\gg r$ holds, where $\lambda$ is the mean free path, in order to realize high-frequency rotation~\cite{Epstein1924,Roth-1990-book,Corson2017,Ahn2018}. We assume diffuse reflection at the surface. Accordingly, the torque induced by the air resistance is calculated as~\cite{Supplement}
\begin{align}
\label{eq:torque}
 \Gamma_{\rm air}=-\frac{8 r^4}{3}\sqrt{\frac{\pi m_{\rm air}}{2k_{\rm B}T}}\Omega_z p,
\end{align}
where $m_{\rm air}$ is the mass of the air molecules, $k_{\rm B}$ is the Boltzmann constant, $T$ is temperature, and $p$ is pressure. In the following estimate, we assume that the particle is in the atmosphere, for which the average molecular mass is $m_{\rm air}=4.78\times 10^{-26}\,{\rm kg}$ and take $r=1\times 10^{-6}\,{\rm m}$ and $T=273\,{\rm K}$~\footnote{In our estimate, a fluctuation of rotation velocity, which is dominantly induced by thermal fluctuation of air resistance torque, becomes much smaller than its average.
Therefore, the noise of rotation frequency does not affect its precious measurement.}. The mean free path satisfies $p \cdot \lambda = k_\mathrm{B} T / (\sqrt{2} \pi \xi^2)$, where $\xi$ is the diameter of an air molecule. Substituting $\xi=3.76\times 10^{-10}\, {\rm m}$, we have $p \cdot \lambda \simeq 6.0 \times 10^{-3}\ \mathrm{N/m}$, and taking $p \lesssim 100\ \mathrm{Pa}$ is sufficient to enter the molecular flow region $\lambda \gg r$.

{\it Gyromagnetic bifurcation.---} First, we set $g_{\rm SR}=1$ in order to see the features of the steady-state rotation. The Euler equation of a spherical particle is presented in Eq.~(\ref{eq:conservation}). Defining the net angular momentum gain as $f(\Omega_z)\equiv \Gamma_{\rm in}(\Omega_z) +\Gamma_{\rm air}(\Omega_z)$, the rotation frequency of the steady state can be obtained by solving $f(\Omega_z)=0$. Our estimate employs the parameters, $M_0=1.557\times 10^5 \, {\rm A}/{\rm m}$, $\alpha = 6.7\times 10^{-5}$, $H_0=2.6\times 10^5\, {\rm A/m}$, of a spin pumping experiment for YIG~\cite{Kajiwara2010}.

Before showing our results, we should explain that both stable and unstable solutions may appear in general. Figure~\ref{fig:geometry} (b) is a schematic graph of $f(\Omega_z)$ and the solutions of $f(\Omega_z)=0$ when $f(\Omega_z)$ has three solutions, $\Omega_{z,i}$ ($i=1,2,3$). Two solutions, $\Omega_{z,1}$ and $\Omega_{z,3}$, are stable because an infinitesimal change in $\Omega_z$ induces a restoring torque. On the other hand, $\Omega_{z,2}$ is an unstable solution because no restoring torque works there. The emergence of three solutions highly depends on the microwave amplitude $h$ and the Gilbert damping parameter $\alpha$, as will be discussed later. Hereafter, we calculate the steady-state rotation frequency as a function of the microwave frequency $\omega$ and study how it changes as $h$ varies. %\del{for fixed $\alpha$}.

%\begin{figure}[tbp]
% \centering
% \includegraphics[width=55mm]{stable.pdf}
% \caption{Net angular momentum gain $f(\Omega_z)$. There are three steady-state solutions, $\Omega_{z,i}$ ($i=1,2,3$). For $\Omega_{z,1}$ and $\Omega_{z,3}$ (red points), an infinitesimal change in $\Omega_z$ produces a restoring torque, and thus, the solutions become stable. In contrast, an infinitesimal deviation subsequently grows for $\Omega_{z,2}$, and thus, that solution is unstable. }
% \label{fig:stable}
%\end{figure}

\begin{figure}[tb]
 \centering
 \includegraphics[width=\linewidth]{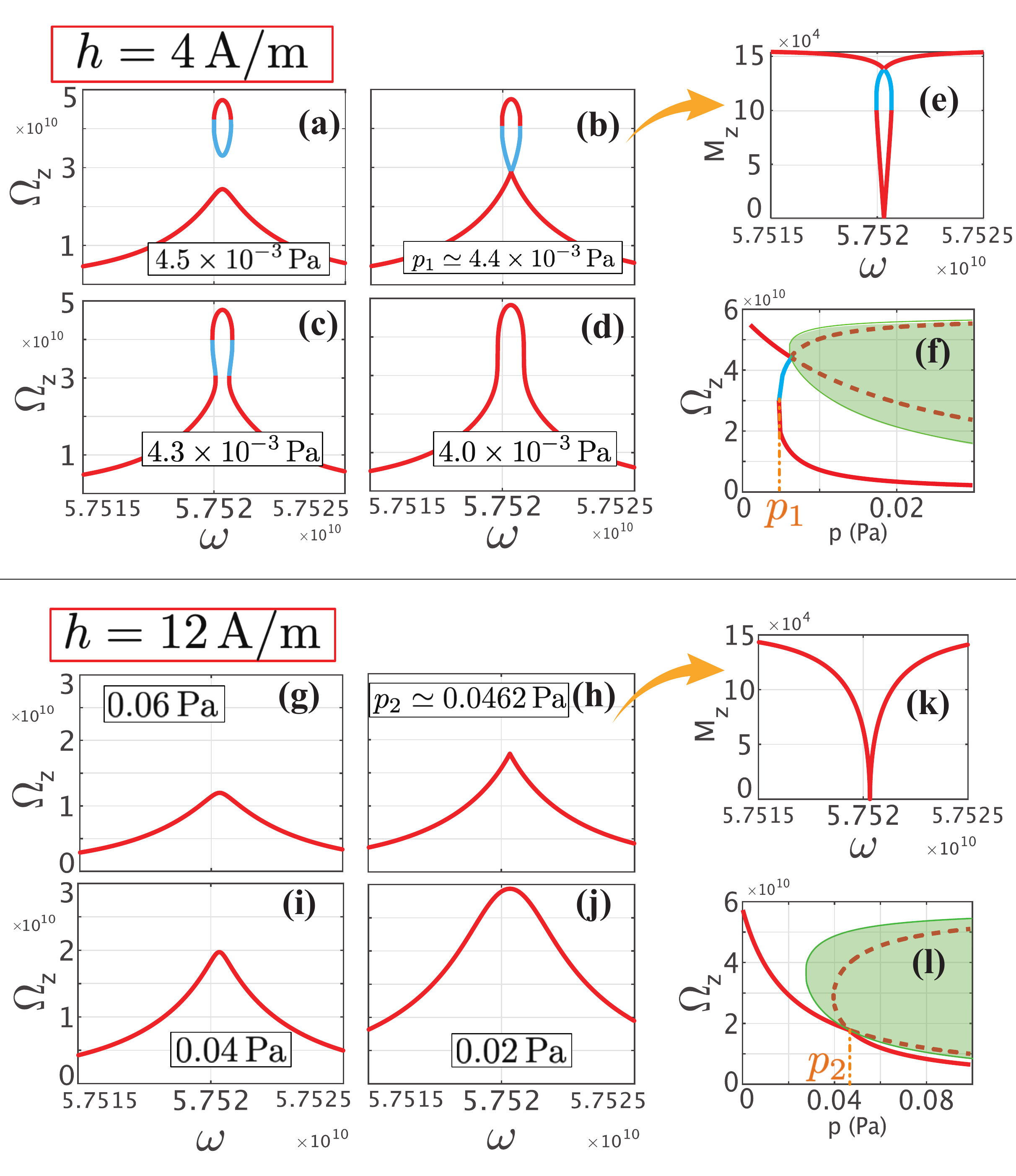}
 \caption{Steady-state rotation frequency $\Omega_z$ for various pressures with microwave amplitudes (a)--(d) $4 \,{\rm A/m}$ and (g)--(j) $12\,{\rm A/m}$. (e)(k) The steady-state solutions of the magnetization $M_z$ that correspond to $\Omega_z$ in (b) and (h), respectively. (f)(l) Red and blue curves express the $p$ dependence of the steady-state solutions for the rotation frequency. The green area is the region in which the $z$ component of the magnetization $M_z$ becomes imaginary. Thus, the physical solutions are the red curves, and the solutions depicted by the blue curves are unstable.) }
 \label{fig:result}
\end{figure}

Figures ~\ref{fig:result}~(a)-(d) show the rotation frequency of the steady state as a function of the microwave frequency for $h=4\,{\rm A/m}$. The red and blue curves indicate the stable and unstable solutions, respectively. For a sufficiently high pressure, only one solution can be realized for arbitrary microwave frequencies (not shown in Fig.~\ref{fig:result}). As the pressure is lowered, two stable solutions and one unstable solution appear (Fig.~\ref{fig:result}~(a)). These new solutions produced by bifurcation appear only near the resonant frequency $\omega_0=-\mu_0\gamma H_0\simeq 57.52\,{\rm GHz}$, while one stable solution exists away from the resonant frequency. As the pressure is lowered, the lower stable branch approaches the upper one and they become connected to each other at a certain pressure, $p_{\rm 1}$ (Fig.~\ref{fig:result}~(b)). Below this critical pressure, the topology of the graph of $\Omega_z$ changes (Fig.~\ref{fig:result}~(c)), and a transition from the lower to the upper branch becomes possible when the microwave frequency $\omega$ approaches $\omega_0$. For a sufficiently low pressure, the bifurcation disappears and only one stable solution exists for any microwave frequency (Fig.~\ref{fig:result}~(d)). In order to see the characteristics of the branches, we show the magnetization of the particle at the critical pressure in Fig.~\ref{fig:result}~(e). The particle has a small (large) magnetization in the upper (lower) branch with a large (small) steady-state rotation frequency. This means that the distribution of the angular momentum between magnetization and rigid-body rotation is different in these two branches.

Figure~\ref{fig:result}~(f) shows the steady-state solutions at $\omega=\omega_0$ in $p$-$\Omega_z$ space. The green area in the figure is the forbidden region in which $M_z$ becomes an imaginary number. The critical pressure is defined as the lowest pressure in the lower branch in Fig.~\ref{fig:result}~(f) and is calculated as~\cite{Supplement}
\begin{align}
\label{eq:pTR}
p_{\rm 1}=\frac{4|\gamma|\mu_0^2M_0h^2}{r\alpha\omega_0^2}\sqrt{\frac{\pi k_\mathrm{B}T}{2 m_{\rm air}}}.
\end{align}
In the present estimate, the critical pressure at $h=4\,{\rm A/m}$ is $p_{\rm 1}\simeq4.4\times10^{-3}\,{\rm Pa}$.

Figures~\ref{fig:result}~(g)-(j) show the steady-state rotation frequency for $h=12\, {\rm A/m}$. In this case, there is no bifurcation and only one stable solution at any frequency and any pressure. Note that the graph of $\Omega_z$ has a cusp at $\omega=\omega_0$ at a specific pressure $p=p_{\rm 2}$. At this pressure, the magnetization reaches zero at $\omega=\omega_0$ (Fig.~\ref{fig:result}~(k)). The bifurcation disappears because the unstable solutions are inside the forbidden region, as shown in Fig.~\ref{fig:result}~(l) which plots the steady-state solutions at $\omega=\omega_0$ on the $p$-$\Omega_z$ plane. The pressure $p=p_{\rm 2}$ corresponds to the crossing point of the two solutions, which is always on the boundary of the forbidden region~\cite{Supplement}. Finally, we should note that for $g_{\rm SR}=1$ the ferromagnetic resonance frequency is not affected by the rotation frequency. This is because the decrease in microwave frequency in the rotating frame (see Eq.~(\ref{eq:H})) is completely compensated by the decrease in the effective magnetic field due to the spin-rotation coupling (the Bernett effect) through the LLG equation~(\ref{LLG}).

Thus, whether the bifurcation occurs or not depends on the pressure. The condition for a bifurcation to appear is obtained by solving $p_{\rm 1} < p_{\rm 2}$ at $\omega=\omega_0$ as $h/H_0 < \alpha/2$~\cite{Supplement}.
%\begin{align}
%\label{eq:condition}
%\frac{h}{H_0} < \frac{\alpha}{2}.
%\end{align}
This indicates the bifurcation disappears at high microwave amplitudes, which is consistent with the results shown in Fig.~\ref{fig:result}.

\begin{figure}[tb]
 \centering
 \includegraphics[width=75mm]{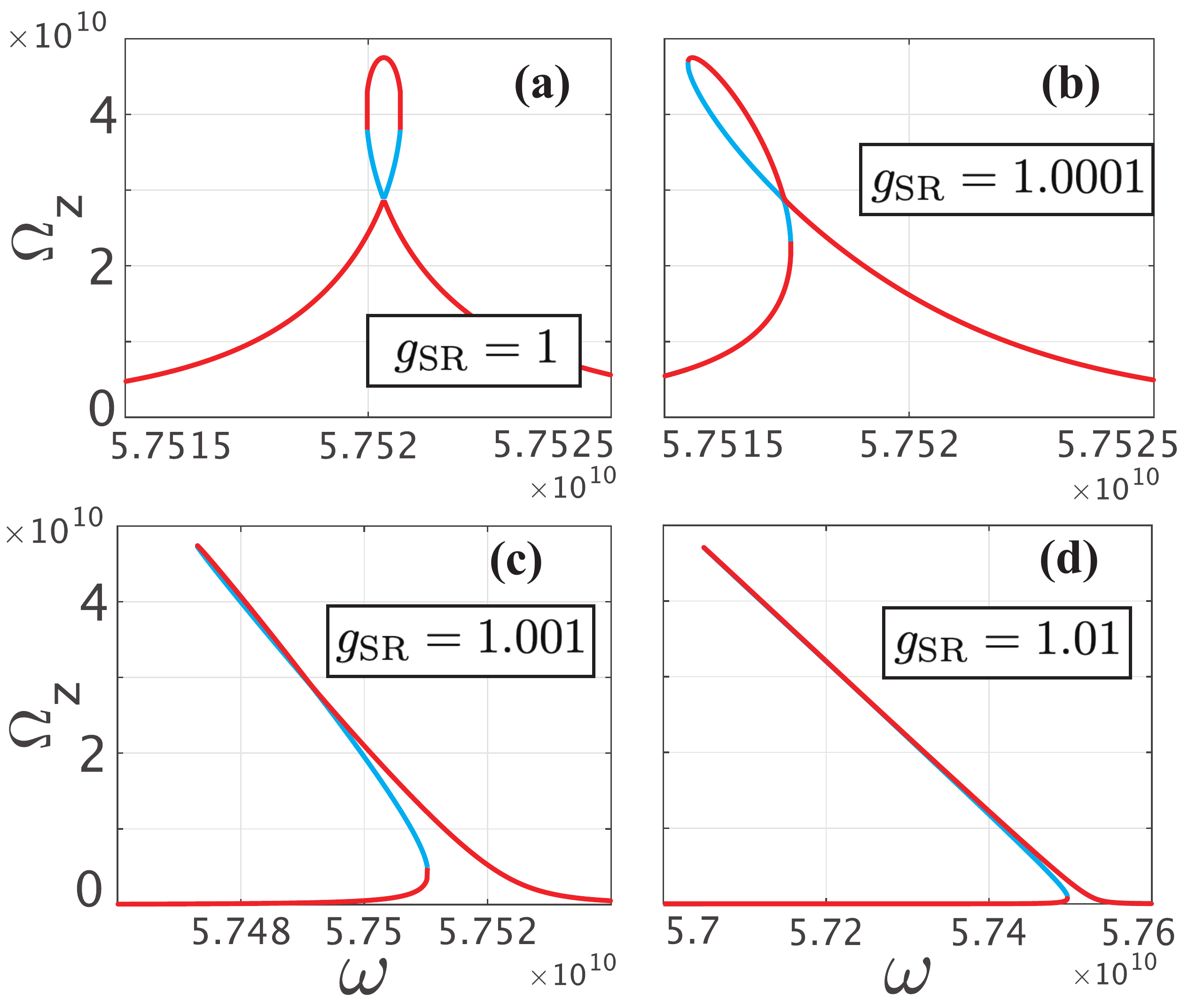}
 \caption{Steady-state solutions for the rotation frequency $\Omega_z$ for various $g_{\rm SR}$: (a) $g_{\rm SR}=1$; (b) $g_{\rm SR}=1.0001$; (c) $g_{\rm SR}=1.001$; (d) $g_{\rm SR}=1.01$. The pressure and microwave amplitude are fixed: $p=p_{\rm 1}$ and $h=4\,{\rm A/m}$. Red (blue) curves are stable (unstable). The scale of the horizontal axis is different in each figure.}
 \label{fig:result2}
\end{figure}

Now, let us consider the case when $g_{\rm SR}$ deviates from 1. Figure~\ref{fig:result2} shows the rotation frequency as a function of the microwave frequency $\omega$ for different values of $g_{\rm SR}$ for $p=p_{\rm 1}$ and $h=4\,{\rm A/m}$. The curves are strongly tilted even for a small deviation in $g_{\rm SR}$ from 1. Furthermore, even when $g_{\rm SR}-1$ is small (see Fig.~\ref{fig:result2}~(c),(d)), the two stable branches exist in a finite range of $\omega$; therefore, transitions from the lower branch to the upper branch always occur as $\omega$ changes, for example $\omega\simeq57.51\, {\rm GHz}$ in Fig.~\ref{fig:result2} (c). These changes originate from incomplete compensation between the microwave frequency shift and the spin-rotation coupling in the rotating frame. This behavior can be utilized for measuring the effective $g$-factor of the spin-rotation coupling. Note that if $g_{\rm SR}-1$ is negative, the graph of $\Omega_z(\omega)$ is tilted in the opposite direction.

%{\it Discussion.---}
% Referee B対策のための段落、先行研究と比較して、詳しくはSupplementで

{\it Conclusion.---} We examined angular momentum transfer between the spins and the mechanical angular momentum of a levitated ferromagnetic particle driven by microwave irradiation in a vacuum. This setup is suitable not only for making precise measurements but also for very fast mechanical rotation, due to the absence of a restoring torque, as shown in recent experiments on levitated optomechanics~\cite{gonzalez2021levitodynamics}. We formulated the steady-state rotation frequency using the LLG equation in combination with angular momentum conservation and estimated it using realistic experimental parameters. We found a bifurcation phenomenon in the solutions for the steady-state rotation frequency when the rotation frequency is very fast and comparable to the microwave frequency. Our setup is a great candidate for satisfying these conditions, which are essential for the strong Barnett effect, in other words, coexistence of the Barnett and the EdH effects. When the $g$-factor for the spin-rotation coupling, $g_{\rm SR}$, is unity, a transition in rotation frequency is observed near the resonant frequency due to transitions between the two branches. Even a slight deviation in the $g$-factor $g_{\rm SR}$ from unity separates the two branches more significantly and the transition in rotation frequency appears in a wider region of pressure. This feature of a bifurcation sensitive to the value of $g_{\rm SR}$ can be used for accurate measurement for the effective $g$-factor.
%最後にひとことダメ押ししたい。
% Refereeの「However, in order to warrant this manuscript for the publication to PRL, I think it needs additional discussions to make this work more important and innovative, which are the methods to verify this phenomena by some experiments. How one can measure accurately the rotation angular frequency and quantify the ``transitions'' between two stable frequencies? 」の、「more important and innovative」に対する明確な回答は？
% 
% spin dynamicsの物理と、optomechanical な物理との「協奏」的な文言？
Our proposal will provide a powerful way to investigate angular momentum conversion from magnetism to macroscopic motion in ferromagnets and will promote interdisciplinary research between two developing research fields, spintronics and optomechanics.
%\blue{Our proposal will pave a new way for studying angular momentum conversion physics in ferromagnets, where both spin and macroscopic mechanical rotation play important roles, and %stimulate further interaction between spintronics and optomechanics. will bridge two developing research fields, spintronics and optomechanics. }

\begin{acknowledgments}
We thank Y. Ominato, H. Taya, and M. Hongo for helpful comments. 
We acknowledge JSPS KAKENHI for Grants (No.~JP20K03831, No.~JP21K03414). 
MM is partially supported by the Priority Program of the Chinese Academy of Sciences, Grant No.~XDB28000000.
D.O.~is supported by the President's PhD Scholarships at Imperial College London, by JSPS Overseas Research Fellowship, by the Institution of Engineering and Technology (IET), and by Funda\c{c}\~ao para a Ci\^encia e a Tecnologia and Instituto de Telecomunica\c{c}\~oes under project UIDB/50008/2020.
T.S.~was supported by the Japan Society for the Promotion of Science through the Program for Leading Graduate Schools (MERIT).
\end{acknowledgments}

\bibliography{./reference}

\end{document}

\title{Supplementary Information: \\ Gyromagnetic bifurcation in a levitated ferromagnetic particle}

\author{T. Sato}
\author{T. Kato}
\affiliation{Institute for Solid State Physics, University of Tokyo, Kashiwa 277-8581, Japan}
\author{Daigo Oue}
\affiliation{Kavli Institute for Theoretical Sciences, University of Chinese Academy of Sciences, Beijing, 100190, China.}
\affiliation{The Blackett Laboratory, Imperial College London, London SW7 2AZ, United Kingdom
}
\affiliation{Instituto Superior T\'{e}cnico, University of Lisbon, 1049-001 Lisboa, Portugal}

\author{M. Matsuo}
\affiliation{Kavli Institute for Theoretical Sciences, University of Chinese Academy of Sciences, Beijing, 100190, China.}
\affiliation{CAS Center for Excellence in Topological Quantum Computation, University of Chinese Academy of Sciences, Beijing 100190, China}
\affiliation{Advanced Science Research Center, Japan Atomic Energy Agency, Tokai, 319-1195, Japan}
\affiliation{RIKEN Center for Emergent Matter Science (CEMS), Wako, Saitama 351-0198, Japan}

\date{\today}

\maketitle
\section{Measurement of the rotation frequency}

The fast rotation of an optically levitated particle has been observed in a number of previous works \cite{Arita2013,Reimann2018,Ahn2018,Monteiro2018}. For example, Ref.~\cite{Reimann2018} reported a GHz rotation of a single $100\, {\rm nm}$ silica particle in vacuum in pressure range from $10^{-3}\,{\rm Pa}$ to $10\,{\rm Pa}$. In Ref.~\cite{Reimann2018}, the authors analyzed the frequency shift of the detection beam which goes through the particle.
Because this shift is independent of the rotation detection mechanism, it has been utilized for accurate measurement of the rotation frequency of the particle.
Though these experiments are mainly performed for non-magnetic particles, levitation of ferromagnetic particles has also been realized recently. 
For example, Ref.~\cite{Huillery2020} achieved the levitation of a spherical iron particle having a diameter around $1\,\mu {\rm m}$ by a Paul trap method. 
Therefore, we expect that both levitation of ferromagnetic particles and measurement of their rotation frequency that are assumed in our work are possible within the present experimental techniques.
We note that our theoretical description based on the LLG equation and the angular-momentum conservation does not depends on the kind of magnetic materials and is applicable to various ferromagnetic particles.

\section{Air Resistance in Vacuum}

We formulate the torque by the air resistance in vacuum referring to Refs.~\cite{Epstein1924,Roth-1990-book,Corson2017,Ahn2018}.
We assume the condition of $\lambda\gg r$, where $\lambda$ is the mean free path of the air and $r$ is the radius of the particle.
We further assume the diffuse reflection at the surface.
Then, the total torque of the air resistance for the particle is given as~\footnote{This equation corresponds to Eq.~(64) in Ref.~\onlinecite{Epstein1924}.}
\begin{align}
\label{eq:torque}
     \Gamma_{\rm air}=-\frac{2\pi r^4\Omega_z}{3}\rho\bar{c},
\end{align}
where $\rho$ and $\bar{c}$ are the density and the mean velocity of molecules, respectively.
From the Maxwell-Boltzmann distribution for an ideal gas, these quantities are given as
\begin{align}
\label{eq:rhoc}
    \rho=nm_{\rm air}=m_{\rm air} \frac{p}{k_{\rm B}T},\,\,\ \bar{c}=\frac{2}{\sqrt{\pi}}\sqrt{\frac{2k_{\rm B}T}{m_{\rm air}}},
\end{align}
where $m_{\rm air}$ is the molecule mass, $n$ is the number of molecules per unit volume, $T$ is the temperature, $p$ is the pressure, and $k_{\rm B}$ is the Boltzmann constant. Substituting Eq.~(\ref{eq:rhoc}) into Eq.~(\ref{eq:torque}), we obtain
\begin{align}
\label{eq:air}
\Gamma_{\rm air}
=-\frac{8 r^4}{3}\sqrt{\frac{\pi m_{\rm air}}{2k_{\rm B}T}}p \Omega_z  \equiv -\beta \Omega_z.
\end{align}
We assume the mass of air molecules, $m=4.78\times 10^{-26}\,{\rm kg}$. We also take $r=10^{-6}\, {\rm m}$ and $T=273\, {\rm K}$.

To check the condition $\lambda\gg r$, we need to evaluate the mean free path, which is given in Ref.~\onlinecite{Roth-1990-book} as
\begin{align}
\lambda=\frac{k_{\rm B}T}{\sqrt{2}\pi\xi^2 p}
\end{align}
where $\xi$ is the diameter of an air molecule.
Taking $\xi=3.76\times 10^{-10}\,{\rm m}$~\footnote{This value for the molecular diameter $\xi$ is taken from Fig.~2.9 of Ref.~\onlinecite{Roth-1990-book}.} and
$T=273\, {\rm K}$, the mean free path of the air satisfies
\begin{align}
  p\cdot\lambda\simeq 6.0\times 10^{-3}  \,\mathrm{N/m}.
\end{align}
The mean free path for several typical pressures is given in Table~\ref{tab:my_label}.
For $r = 10^{-6}\, {\rm m}$, the condition $\lambda \gg r$ is well satisfied when $p \lesssim 100\, {\rm Pa}$.

\begin{table}[tbp]
    \centering
    \begin{tabular}{cc}
        \hline \hline
       $p$ & mean free path $\lambda$  \\\hline
       $10^{-3}\, {\rm Pa}$  & $6\,{\rm m}$\\
       $10^{-2}\, {\rm Pa}$ & $60\,{\rm cm}$\\
       $0.1\, {\rm Pa}$ & $6\,{\rm cm}$\\
       $1\,{\rm Pa}$ & $6\, {\rm mm}$\\
       $10\,{\rm Pa}$ & $0.6 \, {\rm mm}$ \\\hline \hline
    \end{tabular}
    \caption{The pressure dependence of the mean free path for the air molecules.}
    \label{tab:my_label}
\end{table}

\section{Derivation of the angular-momentum conservation law}

We derive the angular-momentum conservation law of the system.
For this purpose, it is convenient to start from the Lagrangian in the rotating frame given as
\begin{align}
{\cal L} =& \frac{1}{2}I\dot{\varphi}^2+\hbar S^{\rm tot}\dot{\phi}_s(\cos\theta_s-1)-{\cal H} \nonumber
\\
=&\frac{1}{2}I\dot{\varphi}^2+\hbar S^{\rm tot}\dot{\phi}_s(\cos\theta_s-1)+\mu_0\gamma {\bm H}\cdot{\bm S}+g_{\rm SR}{\bm \Omega}\cdot{\bm S}\nonumber
\\
=&\frac{1}{2}I\dot{\varphi}^2+\hbar S^{\rm tot}\dot{\phi}_s(\cos\theta_s-1) \nonumber 
\\\nonumber&+\hbar S^{\rm tot}\mu_0\gamma\left\{h\sin\theta_s\left[\cos(\omega t-\varphi)\cos\phi_s\right.\right.
\\\nonumber&\hspace{2cm}\left.\left.+\sin(\omega t-\varphi)\sin\phi_s\right]+H_0\cos\theta_s\right\}
\\
&+g_{\rm SR}\hbar S^{\rm tot}\cos\theta_s\dot{\varphi},
\end{align}
where ${\bm S}^{\rm tot}=S^{\rm tot}(\sin\theta_s\cos\phi_s,\sin\theta_s\sin\phi_s,\cos\theta_s)$, $\varphi$ is a rotation angle whose time derivative corresponds to ${\bm \Omega}=(0,0,\dot{\varphi})$, and we have imposed $\varphi(t=0)=0$ as the initial condition. 
From the Euler-Lagrange equation with respect to $\varphi$, $\dv{}{t}\pdv{{\cal L}}{\dot{\varphi}}=\pdv{{\cal L}}{\varphi}$, we obtain
\begin{align}
\label{eq:conservation}
&\dv{}{t}\left[I\dot{\varphi}+g_{\rm SR}\hbar S^{\rm tot}_z\right] \nonumber
\\&=\hbar S^{\rm tot}\mu_0\gamma h \left[\sin(\omega t-\varphi)\cos\phi_s-\cos(\omega t-\varphi)\sin\phi_s\right] \nonumber
\\
&=\mu_0\gamma\hbar\left[{\bm S}^{\rm tot}\times{\bm H}\right]_z
\equiv\Gamma_{\rm in},
\end{align}
where $\Gamma_{\rm in}=\mu_0\gamma \hbar \left[{\bm S}^{\rm tot}\times{\bm H}\right]_z$ is the torque supplied from the microwave.
By introducing the dissipative function to describe the air resistance, an additional torque $\Gamma_{\rm air}$ is added to the right-hand side of the above equation.
By considering the steady state, $\dot{\varphi}=\Omega_z$, and using $\varphi=\int_0^t\dot{\varphi}dt=\Omega_z t$, we obtain Eq.~(7) in the main text.

\section{Steady-state Euler Equation}

The dynamics of the system is determined by the LLG equation
\begin{align}
\label{eq:LLGEuler}
\dot{{\bm M}} ={\bm M}\times (\mu_0\gamma {\bm H}+g_{\rm SR}{\bm \Omega})+\frac{\alpha}{M}{\bm M}\times \dot{{\bm M}},
\end{align}
and the Euler equation
\begin{align}
\label{eq:Euler}
I\dot{\Omega}_z+g_{\rm SR}\hbar \dot{S}_z^{\rm tot}=\Gamma_{\rm in}+\Gamma_{\rm air},
\end{align}
where we have assumed ${\bm \Omega}=(0,0,\Omega_z)$.
We further assume the $x$ and $y$ components of ${\bm M}$ as
\begin{align}
\label{eq:Msteady}
M_x &= M\cos\left[(\omega -\Omega_z)t+\phi\right], \\\nonumber
M_y &= M\sin\left[(\omega-\Omega_z) t+\phi\right],
\end{align}
where $M^2 = M_0^2 - M_z^2$.
By taking 
\begin{align}
\dot{M_z}=\dot{\Omega_z}=0,
\end{align}
the steady-state solution is obtained as
\begin{align}
\label{eq:phim}
&\tan\phi=\frac{\alpha\frac{M_z}{M_0}(\omega-\Omega_z)}{\omega-\omega_0+\Delta g\Omega_z},\\
&M=
\frac{\mu_0\gamma M_z h}{\sqrt{(\omega-\omega_0+\Delta g\Omega_z)^2+(\alpha(\omega-\Omega_z) M_z/M_0)^2}}.
\label{eq:m}
\end{align}
where $\omega_0=-\mu_0\gamma H_0$ and we defined $\Delta g=g_{\rm SR}-1$. 
Using $M^2+M_z^2=M_0^2$, the equation for $M_z^2$ can be derived. By solving this equation, we obtain
\begin{align}
\label{eq:Mz}
M_z^2&=\frac{1}{2\alpha^2(\omega-\Omega_z)^2/M_0^2} \left[-A + \sqrt{A^2+B}\right], \\
A &= \mu_0^2\gamma^2 h^2+(\omega-\omega_0+\Delta g\Omega_z)^2-\alpha^2(\omega-\Omega_z)^2, \\
B &= 4\alpha^2(\omega-\Omega_z)^2(\omega-\omega_0+\Delta g\Omega_z)^2.
\end{align}
Here, we have chosen the solution in which the sign in front of the square root is plus, 
because $M_z$ has to be real. 
Since $\Gamma_{\rm in}$ is written as
\begin{align}
\Gamma_{\rm in}&=\mu_0\gamma\hbar\left[{\bm S}^{\rm tot}\times{\bm H}\right]_z \nonumber \\
&=-\frac{\alpha V}{M_0\gamma}(\omega-\Omega_z)(M_0^2-M_z^2),
\end{align}
the steady-state Euler equation (\ref{eq:Euler}) becomes
\begin{align}
-\frac{\alpha V}{M_0\gamma}(\omega-\Omega_z)(M_0^2-M_z^2)-\beta\Omega_z=0,
\label{eq:EulerEq}
\end{align}
where $\beta$ is the constant defined in Eq.~(\ref{eq:air}).
Substituting Eq.~(\ref{eq:Mz}) into Eq.~(\ref{eq:EulerEq}), we obtain the cubic equation for $\Omega_z$ as 
\begin{align}
\label{eq:3rd}
& a_3 \Omega_z^3 + a_2 \Omega_z^2 + a_1 \Omega_z + a_0 = 0, 
\end{align}
where
\begin{align}
a_3 &= -\left(\frac{2\beta\alpha}{M_0}\right)^2+\frac{4\beta\alpha}{M_0}\frac{V}{\gamma}(\Delta g^2+\alpha^2), \\
a_2 &= \left(\frac{2\beta\alpha}{M_0}\right)^2\omega+\frac{8\beta\alpha}{M_0}\frac{V}{\gamma}(\Delta g(\omega-\omega_0)-\alpha^2\omega), \\
a_1 &= \frac{4\beta\alpha}{M_0}\frac{V}{\gamma}(\mu_0^2\gamma^2 h^2+(\omega-\omega_0)^2+\alpha^2\omega^2) \nonumber \\
&\hspace{7mm} -4\frac{V^2}{\gamma^2}\mu_0^2\gamma^2h^2\alpha^2, \\
a_0 &= 4\frac{V^2}{\gamma^2}\mu_0^2\gamma^2h^2\alpha^2\omega.
\end{align}
The solution of Eq.~(\ref{eq:3rd}) has to satisfy the condition
\begin{align}
h(\Omega_z) &= \sqrt{A(\Omega_z)^2+B(\Omega_z)} \nonumber \\
&= \mu_0^2 \gamma^2 h^2 + (\omega-\omega_0+\Delta g \Omega_z)^2
+ \alpha^2 (\omega-\Omega_z)^2 \nonumber \\
&\hspace{5mm} + \frac{2\alpha \beta \gamma}{M_0 V}\Omega_z (\omega-\Omega_z) \ge 0,
\label{eq:condition}
\end{align}
which appears when squaring Eq.~(\ref{eq:EulerEq}). The solutions of Eq.~(\ref{eq:3rd}) which don't satisfy Eq.~(\ref{eq:condition}) correspond to negative $M_z^2$, in other words imaginary $M_z$.

\section{Bifurcation and critical pressure}

\begin{figure}
    \centering
    \includegraphics[width=85mm]{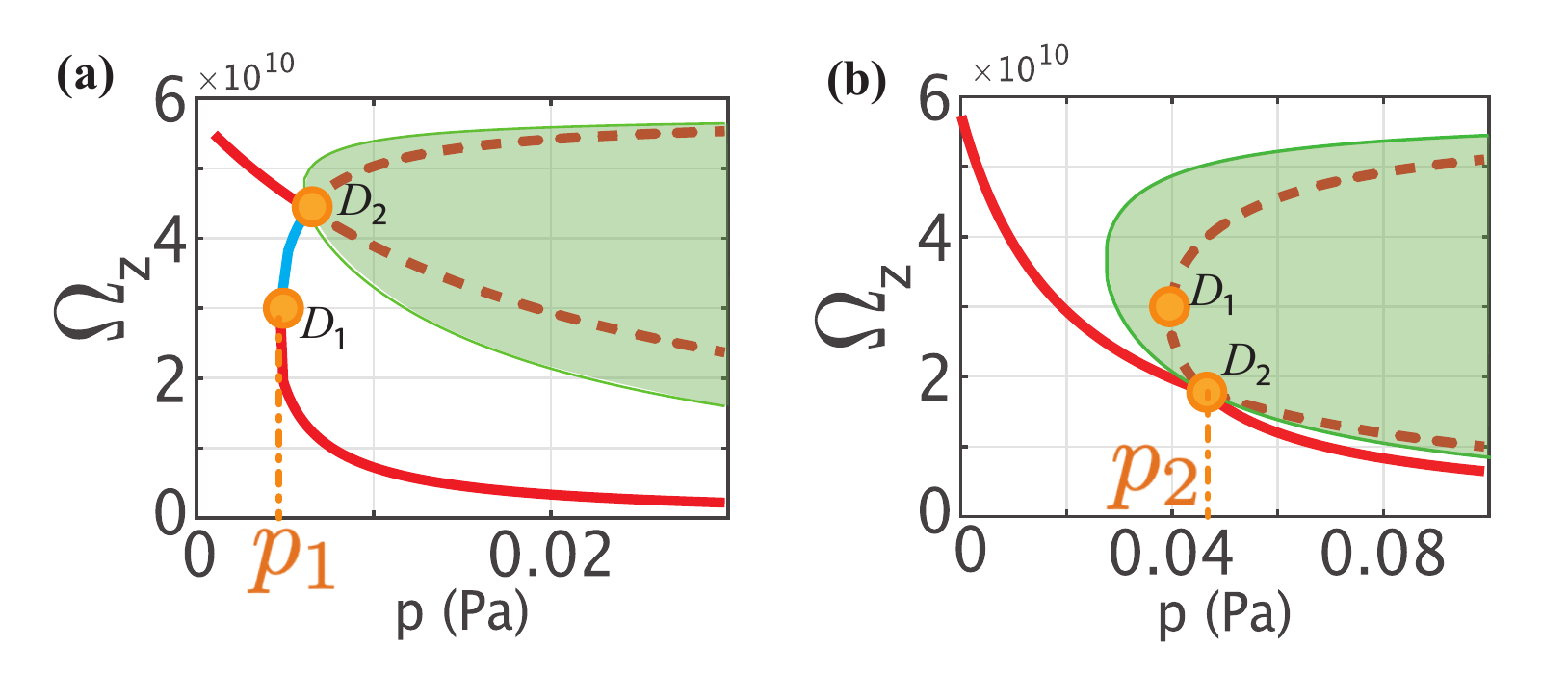}
    \caption{The steady-state rotation frequency $\Omega_z$ as a function of the pressure for $\Delta g=0$ and $\omega=\omega_0$ at (a) $h=4 \, {\rm A/m}$ and (b) $h=12\,{\rm A/m}$. Red (blue) solid curves express the steady-state stable (unstable) solutions of rotation frequency. Green area is the region that the $z$ component of the magnetization $M_z$ becomes imaginary. When solving Eq.~(\ref{eq:3rd}), not only the red and blue solid curves but also red dotted curves appear. However, red dotted curves are prohibited since its $z$ component of the magnetization becomes imaginary.
    }
    \label{fig:d}
\end{figure}

In this section, we describe the condition for bifurcation and detailed calculation on the critical pressure for $g_{\rm SR}=1$ and $\omega=\omega_0$ given in the main text.
We show the real solutions of $\Omega_z$ as a function of the pressure $p$ by the solid lines in Fig.~\ref{fig:d}.
The red (blue) part of the solid line indicates the stable (unstable) solution.
Fig.~\ref{fig:d}~(a) and (b) correspond to the cases of $h=4 \, {\rm A/m}$ and $h=12\,{\rm A/m}$, respectively.
We also show the forbidden region, in which the condition (\ref{eq:condition}) does not holds, by the green hatch in Fig.~\ref{fig:d}.
The dashed lines in the hatch indicate the unphysical solutions, which give imaginary values for $M_z$.

By comparison between Fig.~\ref{fig:d}~(a) and (b), we can see that the behavior of the steady-state solutions is distinct for $h=4 \, {\rm A/m}$ and $h=12\,{\rm A/m}$; one unstable solution exists for weak amplitude of microwave, while it disappears for large amplitude of microwave.
This difference is analyzed by the relative position between the bifurcation point $D_1$ and the crossing point $D_2$ shown in Fig.~\ref{fig:d}.
These two points correspond to the multiple roots of Eq.~(\ref{eq:3rd}) for $\Delta g=0$ and $\omega=\omega_0$.
We define $x$, $y$, and $z$ as
\begin{align}
\label{eq:xyz}
x=\frac{2\beta\alpha}{M_0},\,\,\ y=-\frac{V}{\gamma}\alpha^2,\,\,\ z=-\frac{V}{\gamma}\mu_0^2\gamma^2 h^2\frac{1}{\omega_0^2}.
\end{align}
We note that $x$, $y$, and $z$ are all positive since $\gamma <0$.
Then, the cubic equation given in Eq.~(\ref{eq:3rd}) is rewritten for $\Delta g=0$ and $\omega=\omega_0$ as
\begin{align}
\label{eq:3rd2}
&\Omega_z^3(x+2y)x-\Omega_z^2(x+4y)x\omega_0\\\nonumber&\hspace{1cm}+\Omega_z(4yz+2xz+2xy)\omega_0^2-4yz\omega_0^3=0.
\end{align}
The discriminant of Eq.(\ref{eq:3rd2}) is given as 
\begin{align}
D=4\omega_0^4 x (x-8z)(x^2y-x^2z-4xyz-4y^2z)^2.
\end{align} 
By solving $D=0$, the condition for the multiple roots is given as $x=x_1, x_2$ where
\begin{align}
x_1=8z, \quad x_2 = \frac{2y(z+\sqrt{yz})}{y-z}.
\end{align}
We note that the sign in front of the square root is chosen as a plus since $x$ have to be positive.
The solutions, $x=x_1$ and $x=x_2$, correspond to $D_1$ and $D_2$, respectively.
The steady-state rotational frequencies are calculated for these two solutions as
\begin{align}
&\Omega_z(D_1)=\frac{1}{2}\omega_0, \\
\label{eq:D2}
&\Omega_z(D_2)=\frac{y-z}{y+\sqrt{yz}}\omega_0.
\end{align}

Next, we consider the boundary of the forbidden area (the green hatched area in Fig.~\ref{fig:d}).
It is written with $x$, $y$, and $z$ as
\begin{align}
\label{eq:condition2}
h(\Omega_z)&=-\frac{\gamma}{V} \Bigl[ (x+y)\Omega_z^2+(-x-2y)\omega_0\Omega_z \Bigr. \nonumber \\
&\hspace{12mm} \Bigl. +(y+z)\omega_0^2\Bigr]=0.
\end{align}
We can check that the solution for $D_2$ given by $x=x_2$
and Eq.~(\ref{eq:D2}) satisfies this condition.
This indicates that the crossing point $D_2$ of the steady-state solution is always on the boundary of the forbidden region in Fig.~\ref{fig:d}.
We note that the $D_2$ point disappears when $y<z$; in this case only one branch is observed just like the usual FMR.
For $y>z$, the unstable solution (indicated by the blue line in Fig.~\ref{fig:d}~(a)) exists only when $\Omega_z(D_1)<\Omega_z(D_2)$.
This condition is written as
\begin{align}
\label{eq:h_vs_alpha}
\frac{h}{H_0} < \frac{\alpha}{2}.
\end{align}
When the unstable solution exists, the $D_1$ point corresponds to the critical pressure $p_{\rm 1}$ at which the two branches connect each other:
\begin{align}
\label{eq:pcr}
p_{\rm 1}
= - \frac{2M_0}{\gamma\alpha r}\sqrt{\frac{2\pi k_B T}{ m_{\rm air}}} \left(\frac{h}{H_0}\right)^2.
\end{align}
For $h=4\,{\rm A}/{\rm m}$, which correspond to the situation given in Fig.~3 (b) in main text, the critical pressure is evaluated as $p_{\rm 1}\simeq 4.4\times 10^{-3}\,{\rm Pa}$.

\section{Effect of the Gilbert damping}

\begin{figure}
    \centering
    \includegraphics[width=\linewidth]{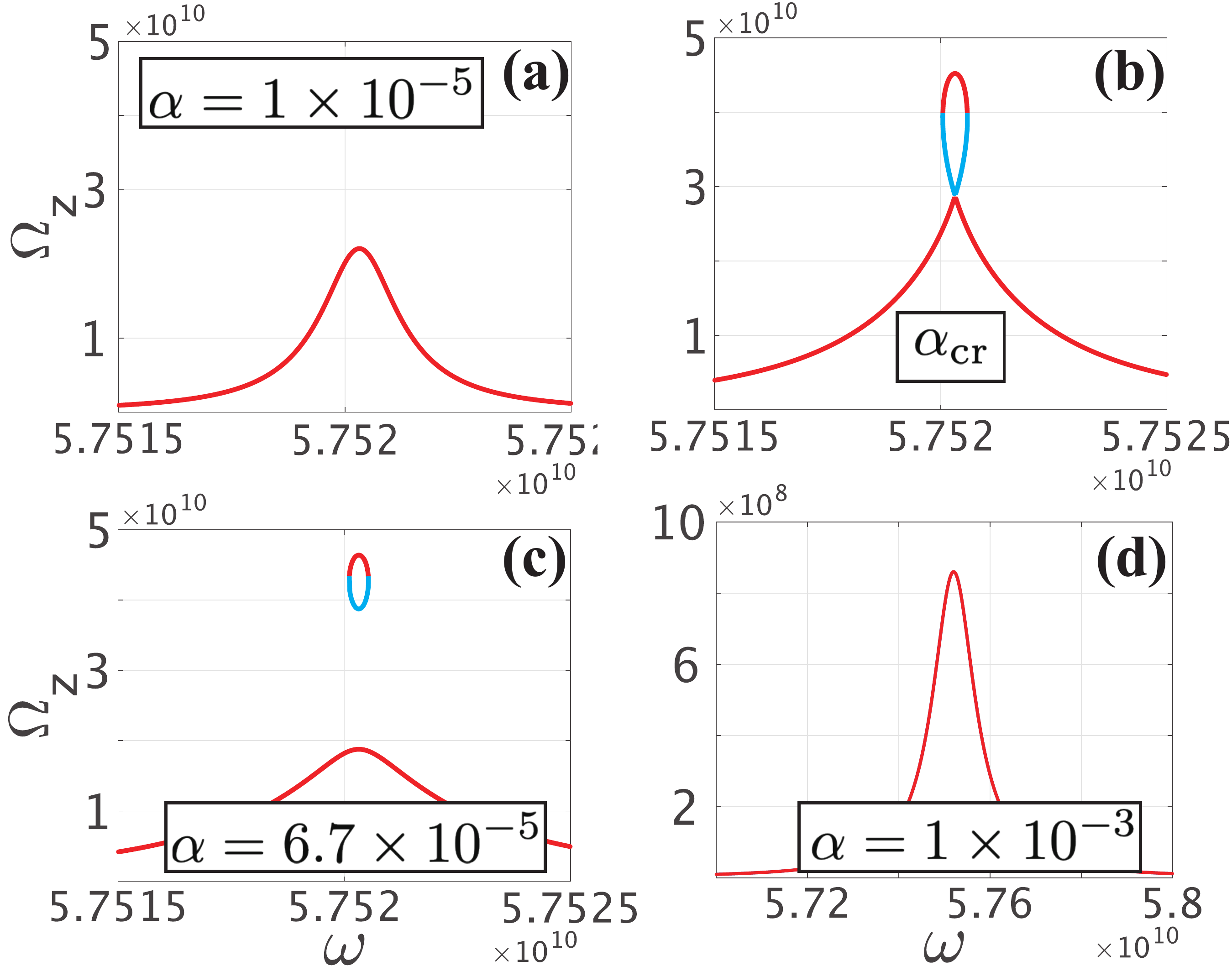}
    \caption{Steady real solutions of rotation frequency for each $\alpha$ at $g_{\rm SR}=1$, $p=5.0\times 10^{-3}\, {\rm Pa}$ and $h=4\,{\rm A/m}$ are described. $\alpha_{\rm cr}\simeq 5.89\times 10^{-5}$ determined from Eq.~(\ref{eq:pcr}).}
    \label{fig:result2}
\end{figure}

We discuss the effect of the Gilbert damping for the steady-state rotation.
We first show the rotation frequency $\Omega_z$ as a function of the microwave frequency $\omega$ for $g_{\rm SR}=1$, $p=5.0\times 10^{-3} \, {\rm Pa}$, and $h=4\,{\rm A/m}$ in Fig.~\ref{fig:result2}.
As the Gilbert damping $\alpha$ is increased from Fig.~\ref{fig:result2}~(a) to (c), the bifurcation first occurs in the lower branch and subsequently the upper branch is disconnected from the lower one. Finally, the pair of the stable and unstable solutions disappears after the bifurcation in the upper branch and only one stable solution is left as shown in (d).
The increase of $\alpha$ gives a similar effect as the decrease of the amplitude of the microwave, $h$.

\begin{figure}[tb]
    \centering
    \includegraphics[width=\linewidth]{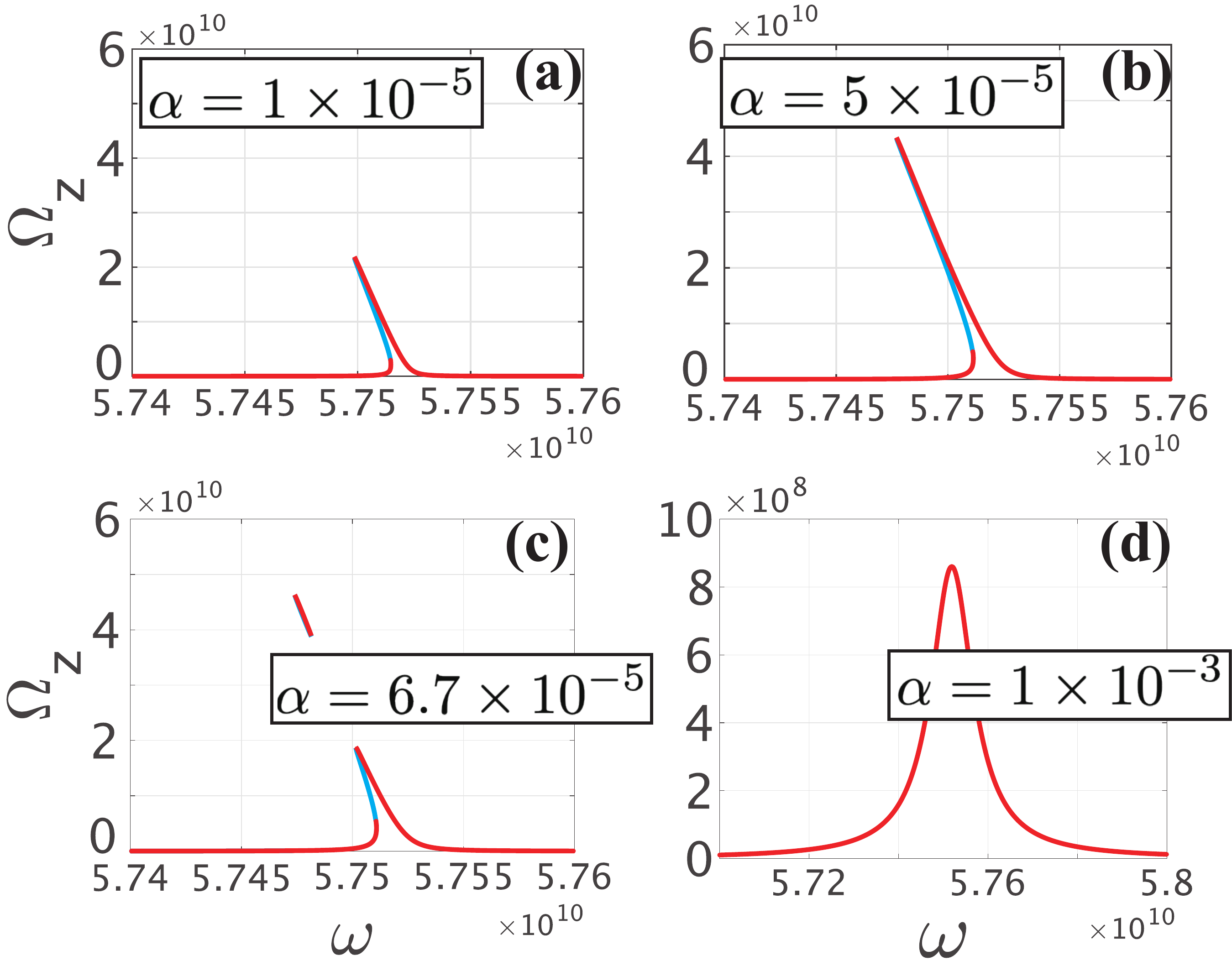}
    \caption{Steady real solutions of rotational frequency $\Omega_z$ for different $\alpha$ are described. $g_{\rm SR}$, $p$, and $h$ is fixed to $1.001$ and $5.0\times 10^{-3}\, {\rm Pa}$, and $4\,{\rm A/m}$ respectively.}
    \label{fig:result3}
\end{figure}

Next, we show the rotation frequency for $g_{\rm SR}=1.001$, $p=5.0\times 10^{-3} \, {\rm Pa}$, and $h=4\,{\rm A/m}$. 
If $\alpha$ is sufficiently small, there are two stable solutions (red lines) and one unstable one (a blue line) as shown in Fig.~\ref{fig:result3}~(a) and (b), and the shape of $\Omega_z(\omega)$ is tilted as pointed out in the main text.
As $\alpha$ increases, the critical pressure $p_{\rm 1}$ becomes lower than the present pressure and the upper branch is disconnected from the lower branch as shown in Fig.~\ref{fig:result3}~(c).
If $\alpha$ further increases, the upper branch disappears and only one solution is realized as shown in Fig.~\ref{fig:result3}~(d), for which the tilt of the graph of $\Omega_z(\omega)$ becomes obscure because of large broadening of the resonant peak.
We note that if the pressure $p$ is much lower, two separated branches exist for any value of the Gilbert damping $\alpha$ since the pressure at the $D_2$ point approaches to $\sim 4.7\times 10^{-3}\, {\rm Pa}$.

In the main text, we chose $\alpha=6.7\times 10^{-5}$ from a typical value for yttrium iron garnet (YIG).
The tilting of the ferromagnetic resonance (FMR) peak for $\Omega_z$ cannot be observed if $\alpha$ becomes large as shown in Fig.~\ref{fig:result3}~(d).
However, the tilting of the FMR peak can be recovered by increasing the microwave amplitude $h$ for such a large value of the Gilbert damping.

\section{$g$-factor for the spin-rotation coupling}

In the main text, we have pointed out the possibility that the $g$-factor for the spin-rotation coupling, $g_{\rm SR}$, deviates from 1. 
Here, we briefly outline how $g_{\rm SR}$ is shifted from 1 for ferromagnetism induced by localized electrons in $3d$ transition-metal ions by considering the crystal field and the spin-orbit coupling following Ref.~\onlinecite{White-2007-book}.

We consider a localized spin in an ion of $3d$ transition metals whose Hamiltonian is given as
\begin{align}
H=H_{\rm intra}+H_{\rm cry}+H_{\rm SO}+H_{\rm Z}+H_{\rm SR}.
\end{align}
Here, $H_{\rm intra}$ and $H_{\rm cry}$ describe the intraatomic Coulomb and the crystal field, respectively.
The last three terms are the Hamiltonians of the spin-orbit coupling, the Zeeman energy, and the spin-rotation coupling and are given as
\begin{align}
& H_{\rm SO} = \lambda {\bm L}\cdot{\bm S}, \\
& H_{\rm Z} = \mu_B{\bm H}\cdot({\bm L}+2{\bm S}), \\
& H_{\rm SR} = {\bm \Omega}\cdot({\bm L}+{\bm S}),
\end{align}
respectively.
We assume $H_0 = H_{\rm intra}+ H_{\rm cry}\gg H_{\rm SO}, H_{\rm Z}, H_{\rm SR}$, which is satisfied for $3d$ transition-metal ions.
We assume that the ground state of $H_0$ is nondegenerate and express it as $\ket{G,M_s}$, where $M_s$ is the spin quantum number. 
Since $H_0$ does not mix spin and orbit states, this eigenfunctions are written as the products, $\ket{G,M_s}=\ket{G}\ket{M_s}$.
Using the perturbation theory, the first- and the second-order corrections to the energy for an orbitally nondegenerate ground state, $E^{(1)}, E^{(2)}$, are derived as
\begin{align}
E^{(1)} 
&=2\mu_B{\bm H}\cdot{\bm S}+{\bm \Omega}\cdot{\bm S},\\
E^{(2)}
&=-\sum_{i\neq G}\frac{\left|\bra{G}{\bm L}\cdot(\lambda{\bm S}+\mu_B{\bm H}+{\bm \Omega})\ket{i}\right|^2}{E_i^{(0)}-E_G^{(0)}}
\end{align}
where we have used the fact that $\bra{G}{\bm L}\ket{G}$ vanish. When we define
\begin{align}
    \Lambda_{\mu\nu}=\sum_{i\neq G}\frac{\bra{G}L_\mu\ket{i}\bra{i}L_\nu\ket{G}}{E_i^{(0)}-E_G^{(0)}},
\end{align}
the energy correction reduces to 
\begin{align}
E^{(2)}=-\Lambda_{\mu\nu}&\left[\lambda^2 S_\mu S_\nu + (\mu_B H_\mu+\Omega_\mu) (\mu_B H_\nu+\Omega_\nu)\right.\nonumber
\\
&\left.+2\lambda S_\mu (\mu_B H_\nu +\Omega_\nu)\right]
\end{align}
where the Einstein summation convention has been used. By combining these results, the effective Hamiltonian becomes
\begin{align}
H_{\rm eff}&=H_0+g^Z_{\mu\nu}\mu_B H_\mu S_\nu+g^{\rm SR}_{\mu\nu}\Omega_\mu S_\nu-\lambda^2\Lambda_{\mu\nu}S_\mu S_\nu 
\nonumber\\&\hspace{5mm} -\Lambda_{\mu\nu} (\mu_B H_\mu+\Omega_\mu) (\mu_B H_\nu+\Omega_\nu),\\
g^Z_{\mu\nu} &=2\delta_{\mu\nu}-2\lambda\Lambda_{\mu\nu}, \\
g^{\rm SR}_{\mu\nu}&=\delta_{\mu\nu}-2\lambda\Lambda_{\mu\nu}.
\end{align}
where $g^Z_{\mu\nu}$ and $g^{\rm SR}_{\mu\nu}$ are $g$-factor tensors for the Zeeman term and the spin-rotation coupling, respectively.
This result indicates that the spin-orbit interaction changes not only the $g$-factor of the Zeeman term but also that of the spin-rotation coupling in the same way through $\Lambda_{\mu\nu}$.
In general, $\Lambda_{\mu\nu}$ can be diagonalized by taking a certain principle axis. 
If the crystal field has cubic symmetry, $\Lambda_{\mu\nu}$ becomes
\begin{align}
\Lambda_{\mu\nu}=0\, (\mu\neq\nu),\,\,\, \Lambda_{xx}=\Lambda_{yy}=\Lambda_{zz}.
\end{align}
This means an isotropic constant shift of the $g$-factor. 
In the main text, we have assumed an isotropic change of the $g$-factor for simplicity.

\section{Stability analysis of the rotation}

Through our article, we have assumed that the in-plane rotation velocities, $\Omega_x$ and $\Omega_y$, are zero. 
We have checked numerically that the in-plane rotation always decays even when the initial condition includes the in-plane rotation.
This indicates that the uniaxial rotation around the $z$ axis is stable against perturbation to incline the rotation axis.
In this section, we show the stability of the uniaxial rotation by  solving Eqs.~(\ref{eq:body}) and (\ref{eq:Eulereq}) within a simple approximation, which leads to the same result obtained by numerical calculation.
%In this section, we will show that this assumption is justified in a usual optical levitation setup.
%From now on, we analytically solve Eq.~(\ref{eq:body})(\ref{eq:Eulereq}) by using some assumptions. The results obtained by numerical way and analytical way are qualitatively consistent.

In order to discuss a general formalism of our model, we have to ride on the body frame in which the in-plane rotation is also allowed.
The coordinate transformation from the laboratory frame, ${\bm e}_x,\,{\bm e}_y,\,{\bm e}_z$, to the body frame, ${\bm e}_{x'},\,{\bm e}_{y'},\,{\bm e}_{z'}$, can be implemented by using Euler angles, $(\varphi,\theta,\psi)$. The transformation matrix ${\rm R}$ (${\bm A}_b={\rm R}{\bm A}$, where ${\bm A}$ is a vector in the lab frame, and ${\bm A}_b$ is that in the body frame) is given as
\begin{align}
{\rm R}&=\begin{pmatrix}
\cos\psi & \sin\psi & 0 \\ -\sin\psi & \cos\psi & 0 \\ 0&0&1
\end{pmatrix}
\cdot 
\begin{pmatrix}
1&0&0\\ 0&\cos\theta & \sin\theta \\ 0& -\sin\theta & \cos\theta
\end{pmatrix}
\\\nonumber
&\hspace{2cm}\cdot
\begin{pmatrix}
\cos\varphi & \sin\varphi & 0 \\ -\sin\varphi & \cos\varphi & 0 \\ 0 & 0& 1
\end{pmatrix}
.
\end{align}
The rotation frequency ${\bm \Omega}$ is expressed by Euler angles as
\begin{align}
\label{eq:Omega}
\begin{cases}
\Omega_{x'}=\dot{\varphi}\sin\theta\sin\psi +\dot{\theta}\cos\psi,
\\
\Omega_{y'}=\dot{\varphi}\sin\theta\cos\psi - \dot{\theta}\sin\psi,
\\
\Omega_{z'}=\dot{\varphi}\cos\theta + \dot{\psi}.
\end{cases}
\end{align}
When $\theta=\varphi=0,\, \dot{\theta}=\dot{\varphi}=0$, the body frame rotates only around $z$ axis, and it becomes the same as the model of our article. For stability analysis, it is sufficient to consider states of $\theta\ll 1$.
%Hereafter, we focus on $\theta\ll 1$ since it is enough for the purpose of stability analysis. 
The dynamics of the magnetization in the body frame, ${\bm M}_b$, is determined from a modified version of Eq.~(\ref{eq:LLGEuler}):
\begin{align}
\label{eq:body}
\dot{{\bm M}}_b ={\bm M}_b\times (\mu_0\gamma {\bm H}_b+{\bm \Omega}_b)+\frac{\alpha}{M}{\bm M}_b\times \dot{{\bm M}}_b,
\end{align}
where $g_{\rm SR}$ is set as unity for simplicity and ${\bm H}_b$ is defined as
\begin{align}
{\bm H}_b&
={\rm R}{\bm H}={\rm R}\begin{pmatrix}
    h\cos\omega t\\h\sin\omega t\\H_0\\
\end{pmatrix}
\\\nonumber & \simeq \begin{pmatrix}
h\cos(\omega t-\varphi-\psi) \\ h\sin (\omega t-\varphi-\psi) \\ H_0
\end{pmatrix}
+\theta\begin{pmatrix}
H_0\sin\psi \\ H_0\cos\psi \\ -h\sin(\omega t-\varphi)
\end{pmatrix}.
\end{align}
By expanding ${\bm M}_b$ with respect to $\theta$ as
\begin{align}
{\bm M}_b = {\bm M} + \theta {\bm M}_1 +{\cal O}(\theta^2),
\end{align}
we can derive equations from the zeroth- and first-order of $\theta$.
The zeroth-order equation leads to the equation of ${\bm M}$ given in the main text and therefore the steady-state solution of ${\bm M}$ is given as
\begin{align}
{\bm M} 
\equiv
\begin{pmatrix}
M_{x'}\\M_{y'}\\M_{z'}
\end{pmatrix}
= \begin{pmatrix}M\cos[(\omega-\Omega_{z'})t+\phi] \\ M\sin[(\omega-\Omega_{z'})t+\phi] \\ M_z
\end{pmatrix}.
\end{align}
Here, we used $\Omega_{z'}=\dot{\varphi}\cos\theta+\dot{\psi}\simeq \dot{\varphi}+\dot{\psi}$ and assumed that $\dot{\varphi}$ and $\dot{\psi}$ are almost constant.
On the other hand, from the equation of the first order of $\theta$, ${\bm M}_1\equiv (M_{1x'}, M_{1y'}, M_{1z'})$ obeys
\begin{widetext}
\begin{align}
\nonumber
&\begin{pmatrix}
\partial_t & \frac{\alpha}{M_0}M_{z'}\partial_t-\mu_0\gamma H_0-\Omega_{z'} & -\frac{\alpha}{M_0}M_{y'}\partial_t+\mu_0\gamma H_{y'} +\frac{\alpha}{M_0}\dot{M}_{y'}\\ -\frac{\alpha}{M_0}M_{z'}\partial_t+\mu_0\gamma H_0+\Omega_{z'} & \partial_t & \frac{\alpha}{M_0}\partial_t-\mu_0\gamma H_{x'}-\frac{\alpha}{M_0}\dot{M}_{x'} \\ \frac{\alpha}{M_0}M_{y'}\partial_t-\mu_0\gamma H_{y'} -\frac{\alpha}{M_0}\dot{M}_{y'} & -\frac{\alpha}{M_0}M_{x'}\partial_t+\mu_0\gamma H_{x'}-\frac{\alpha}{M_0}\dot{M}_{x'} & \partial_t
\end{pmatrix}
\begin{pmatrix}
M_{1x'}\\M_{1y'}\\M_{1z'}
\end{pmatrix}
\\ &\hspace{8cm}
=-\mu_0\gamma 
\begin{pmatrix}
 H_0 M_{z'} \cos\psi + h\sin(\omega t-\varphi) M_{y'}  \\ - H_0 M_{z'} \sin\psi -h\sin(\omega t-\varphi) M_{x'} \\ H_0 (\sin\psi M_{y'} - \cos\psi M_{y'})
\end{pmatrix},
\end{align}
\end{widetext}
where $(H_{x'},H_{y'})\equiv h(\cos(\omega-\Omega_{z'})t, \sin(\omega-\Omega_{z'})t)$. Extracting upper two equations, and assuming $h\ll H_0$, these equations are reduced to 
\begin{widetext}
\begin{align}
\nonumber
&\begin{pmatrix}
\partial_t - i\dot{\psi} +i \left(-\frac{\alpha}{M_0}M_{z'}(\partial_t-i\dot{\psi})+\mu_0\gamma H_0 +\Omega_{z'}\right) & i\frac{\alpha}{M_0}M_{+}\partial_t-i\mu_0\gamma H_+ + \frac{\alpha}{M_0}(\omega-\dot{\varphi}-\dot{\psi})M_{+}
\end{pmatrix}
\begin{pmatrix}
M_1 \\ M_{1z'}
\end{pmatrix}
\\&\hspace{8cm}=-\mu_0\gamma H_0 M_{z'},
\label{eq:stability}
\end{align}
\end{widetext}
where $M_1\equiv (M_{1x'}+i M_{1y'})e^{i\psi}$, $M_{+}\equiv M e^{i(\omega t-\varphi +\phi)}$ and $H_{+}\equiv (H_{x'}+ i H_{y'})e^{i\psi}=h e^{i(\omega t-\varphi)}$. $M_{+}$ and $H_+$ oscillate with the microwave frequency $\omega$ because $\dot{\varphi}$ is much smaller than $\omega$. Since the dynamics of $M_1$ is roughly dominated by slowly oscillating terms, we can drop these fast oscillating terms. Technically speaking, $M_{+}$ and $H_+$ becomes zero by being integrated out from $t=0$ to $t=2\pi/\omega$. Then, Eq.~(\ref{eq:stability}) becomes
\begin{align}
&\left(1-i\frac{\alpha}{M_0}M_{z'}\right)\partial_t M_1 \\\nonumber&+\left[i(\mu_0\gamma H_0+\Omega_{z'}-\dot{\psi})-\frac{\alpha}{M_0}M_{z'}\dot{\psi}\right]M_1=-\mu_0\gamma M_{z'},
\end{align}
and the steady-state solution of this is obtained as
\begin{align}
\label{eq:bodyLLG}
M_1=\frac{-\mu_0\gamma H_0M_{z'}}{i(\mu_0\gamma H_0+\Omega_{z'}-\dot{\psi})-\frac{\alpha}{M_0}M_{z'}\dot{\psi}}.
\end{align}

Euler equation of the particle is written as 
\begin{align}
\label{eq:Eulereq}
\begin{cases}
I_1\dot{\Omega}_{x'} + \Delta I \Omega_{z'}\Omega_{y'}=\left[{\bm \Gamma}_g+{\bm \Gamma}_{\rm optical}+{\bm \Gamma}_{\rm air}\right]_{x'},
\\
I_1\dot{\Omega}_{y'} - \Delta I \Omega_{z'}\Omega_{x'}= \left[{\bm \Gamma}_g+{\bm \Gamma}_{\rm optical}+{\bm \Gamma}_{\rm air}\right]_{y'},
\\
I_2\dot{\Omega}_{z'}=\left[{\bm \Gamma}_g+{\bm \Gamma}_{\rm optical}+{\bm \Gamma}_{\rm air}\right]_{z'},
\end{cases}
\end{align}
where we assumed that the particle is symmetrical top and $I_1$ ($I_2$) is a principle moment of inertia around $x',\,y'$ ($z'$) axis and $\Delta I\equiv I_2-I_1$.
Here, ${\bm \Gamma}_{g}$ is a torque generated from Gilbert damping given as
\begin{align}
{\bm \Gamma}_g = -\frac{V}{\gamma}\frac{\alpha}{M_0} {\bm M}_b \times \dot{{\bm M}}_b,
\end{align}
${\bm \Gamma}_{\rm optical}$ is a torque from optical trap, and ${\bm \Gamma}_{\rm air}=-\beta{\bm \Omega}$ is a torque from air resistance.
Using Eq.~(\ref{eq:Omega}), equation of motion for $\theta$, $\varphi$ and $\psi$ is written as
%the dynamics of $\theta$ and $\varphi$ is obtained as
\begin{align}
\label{eq:theta}
&I_1\ddot{\theta}+I_1\dot{\varphi}\dot{\psi}\sin\theta+\Delta I \Omega_{z'}\dot{\varphi}\sin\theta \nonumber \\
&=  \left[{\bm \Gamma}_g+{\bm \Gamma}_{\rm optical}+{\bm \Gamma}_{\rm air}\right]_{x'} \cos\psi \nonumber\\
&\hspace{2cm}- \left[{\bm \Gamma}_g+{\bm \Gamma}_{\rm optical}+{\bm \Gamma}_{\rm air}\right]_{y'} \sin\psi,\\
\label{eq:varphi}
&I_1(\ddot{\varphi}\sin\theta+\dot{\varphi}\dot{\theta}\cos\theta-\dot{\theta}\dot{\psi})-\Delta I \Omega_{z'}\dot{\theta} \nonumber \\
&=  \left[{\bm \Gamma}_g+{\bm \Gamma}_{\rm optical}+{\bm \Gamma}_{\rm air}\right]_{x'} \sin\psi \nonumber \\
&\hspace{2cm}+ \left[{\bm \Gamma}_g+{\bm \Gamma}_{\rm optical}+{\bm \Gamma}_{\rm air}\right]_{y'} \cos\psi,
\\
&
I_2(\ddot{\psi}+\ddot{\varphi}\cos\theta-\dot{\varphi}\dot{\theta}\sin\theta)=\left[{\bm \Gamma}_g+{\bm \Gamma}_{\rm optical}+{\bm \Gamma}_{\rm air}\right]_{z'}.
\end{align}
Neglecting the fast oscillating term, the first term in the bracket of the right hand side is approximated as
\begin{align}
&\Gamma_{g,x'}\cos\psi - \Gamma_{g,y'}\sin\psi \simeq -\frac{V}{\gamma}\frac{\alpha}{M_0}\theta\dot{\psi}M_{z'}\Re(M_1),\\
&\Gamma_{g,x'}\sin\psi + \Gamma_{g,y'}\cos\psi \simeq -\frac{V}{\gamma}\frac{\alpha}{M_0}\theta\dot{\psi}M_{z'}\Im(M_1),
\\
&\Gamma_{g,z'}\simeq -\frac{V}{\gamma}\frac{\alpha}{M_0}(\omega-\Omega_{z'})(M_0^2-M_{z'}^2) .
\end{align}
The torque from optical trap, ${\bm \Gamma}_{\rm optical}$, 
 which usually works to restore the axis of inertia to $z$ axis, is written as
%The torque from optical trap, ${\bm \Gamma}_{\rm optical}$, should work to restore axis of inertia to $z$ axis, thus it should be given as
\begin{align}
&\Gamma_{{\rm optical},x'}\cos\psi - \Gamma_{{\rm optical},y'}\sin\psi = -\xi \theta, \\
&\Gamma_{{\rm optical},x'}\sin\psi + \Gamma_{{\rm optical},y'}\cos\psi = 0, 
\\
&\Gamma_{{\rm optical},z'}=0,
\end{align}
where $\xi$ is a coefficient of the optical torque. 
In addition, we take the spherical particle limit $\Delta I=I_2-I_1 \rightarrow 0$ as considered in the main text.
By these assumptions, Eq.~(\ref{eq:theta}) and (\ref{eq:varphi}) are simplified as
\begin{align}
\label{eq:theta2}
& I_1\ddot{\theta} + \beta\dot{\theta} + \left(I_1\dot{\varphi}\dot{\psi}+\xi + \frac{V}{\gamma}\frac{\alpha}{M_0}\dot{\psi}M_{z'}\Re(M_1) \right)\theta \simeq  0, \\
\label{eq:varphi2}
& I_1\theta\ddot{\varphi} + (I_1\dot{\theta}+\beta\theta)\dot{\varphi} - I_1\dot{\theta}\dot{\psi} + \frac{V}{\gamma}\frac{\alpha}{M_0}\theta\dot{\psi}M_{z'}\Im(M_1)  \simeq  0, \\
\label{eq:psi2}
& I_1(\ddot{\psi}+\ddot{\varphi})\simeq -\beta\Omega_{z'}-\frac{V}{\gamma}\frac{\alpha}{M_0}(\omega-\Omega_{z'})(M_0^2-M_{z'}^2).
\end{align}
In our setup, we estimate $I_1\simeq 9\times 10^{-27}\,{\rm N\cdot m\cdot s^2}$, $\beta\simeq 5\times 10^{-29}\,{\rm N\cdot m \cdot s}$ and 
\begin{align}
\frac{V}{\gamma}\frac{\alpha}{M_0}M_{z'}\Im(M_1)&\simeq \frac{V}{\gamma}\alpha M_0 \frac{\omega_0-\dot{\varphi}}{\omega_0} \nonumber \\
&\hspace{-10mm} \sim (\omega_0-\dot{\varphi})\times (-4.3\times 10^{-39})\,{\rm N\cdot m\cdot s}, \\
\frac{V}{\gamma}\frac{\alpha}{M_0}M_{z'}\Re(M_1) &\simeq -\frac{V}{\gamma}\alpha^2 M_{0} \frac{\dot{\psi}}{\omega_0} \nonumber \\ 
&\hspace{-10mm} \sim \dot{\psi}\times 2.9\times 10^{-43}\, {\rm N\cdot m\cdot s}.
\end{align}
In addition, since the optical force is typically order of sub-pico Newton, the torque $\xi$ on a particle with radius $1\, \mu{\rm m}$ from optical trap is estimated to be $\xi \sim 10^{-15} \, {\rm N}\cdot{\rm m}.$ 
Considering that $\dot{\psi}$ is order of $10\,{\rm GHz}$ at most, the third term in the bracket of Eq.~(\ref{eq:theta2}) can be neglected.

The right hand side of Eq.~(\ref{eq:psi2}) is same as the total torque $f(\Omega_z)$ defined in the main text, and vanishes in the stable solutions of $\Omega_z$, which indicates
\begin{align}
\Omega_{z'}\simeq \dot{\psi}+\dot{\varphi}= {\rm const.} 
\end{align} 
Substituting $\dot{\psi}=\Omega_{z'}-\dot{\varphi}$, and treating $\Omega_{z'}$ as constant value, Eq.~(\ref{eq:theta2})(\ref{eq:varphi2}) are transformed to
\begin{align}
\label{eq:theta3}
& I_1\ddot{\theta} + \beta\dot{\theta} + \left(I_1\dot{\varphi}\dot{\psi}+\xi\right)\theta \simeq  0, \\
\label{eq:varphi3}
\nonumber
&I_1\theta\ddot{\varphi} + \left(2I_1\dot{\theta}+\beta\theta-\frac{V}{\gamma}\alpha
M_0 \theta \left(1+\frac{\Omega_{z'}}{\omega_0}\right)\right)\dot{\varphi}  
\\
&\hspace{1cm}\simeq  \left(I_1\dot{\theta}-\frac{V}{\gamma}\alpha M_0 \theta \right)\Omega_{z'},
\end{align}
up to the first order of $\dot{\varphi}$. 
From Eq.~(\ref{eq:theta3}), $\dot{\varphi}$ can be expressed as
%as a function of $\theta$ %When $\dot{\varphi}\ll \Omega_{z'}$ as we assumed before, $\dot{\varphi}$ can be expressed as a function of $\theta$ from Eq.~(\ref{eq:theta3});
\begin{align}
\label{eq:varphi4}
\dot{\varphi}=-\frac{1}{I_1\Omega_{z'}\theta}\left(I_1\ddot{\theta}+\beta\dot{\theta}+\xi\theta\right).
\end{align}
By substituting Eq.~(\ref{eq:varphi4}) into Eq.~(\ref{eq:varphi3}), we obtain
\begin{align}
\label{eq:theta4}
&I_1\dddot{\theta}+(2\beta+A)\ddot{\theta} \nonumber\\
&\hspace{5mm}+\left(\frac{\beta(\beta+A)}{I_1}+2\xi +I_1\Omega_{z'}^2+\beta\frac{\dot{\theta}}{\theta}+I_1\frac{\ddot{\theta}}{\theta}\right)\dot{\theta} \nonumber \\
& \hspace{5mm}+\left(\frac{\xi(\beta+A)}{I_1}+B\Omega_{z'}^2\right)\theta=0,
\\
& A=-\frac{V}{\gamma}\alpha M_0 \left(1+\frac{\Omega_{z'}}{\omega_0}\right) \ (>0) , \\ 
& B=-\frac{V}{\gamma}\alpha M_0 \ (>0).
\end{align}
We assume that the non-linear terms proportional to $\dot{\theta}^2/\theta$ and $\ddot{\theta}\dot{\theta}/\theta$ in Eq.~(\ref{eq:theta4}) can be neglected (the validity of this assumption will be checked later).
Then, Eq.~(\ref{eq:theta4}) is reduced to a linear homogeneous differential equation of third order;
%When we neglect the non-linear terms in Eq.~(\ref{eq:theta4}), whose validity is checked later, Eq.~(\ref{eq:theta4}) is reduced to a linear homogeneous differential equation of third order;
\begin{align}
\label{eq:theta5}
&\dddot{\theta}+a_2\ddot{\theta}+a_1\dot{\theta}+a_0\theta=0,
\\\nonumber & a_2=(2\beta+A)/I_1 \, (>0),
\\\nonumber & a_1=\left(\frac{\beta(\beta+A)}{I_1}+2\xi +I_1\Omega_{z'}^2\right)/I_1 \, (>0),
\\\nonumber & a_0=\left(\frac{\xi(\beta+A)}{I_1}+B\Omega_{z'}^2\right)/I_1 \, (>0).
\end{align}
According to Routh–Hurwitz stability criterion, the general solution of Eq.~(\ref{eq:theta5}) decays in time
%always attenuates 
when and only when $a_2,a_1,a_0>0$ and $a_2a_1>a_0$. 
Using %the estimations; 
$I_1\simeq 9\times 10^{-27}\,{\rm N\cdot m\cdot s^2}$, $\beta\simeq 5\times 10^{-29}\,{\rm N\cdot m\cdot s}$, $\xi\simeq 10^{-15}\,{\rm N\cdot m}$, $B\simeq 2.5\times 10^{-28}\,{\rm N\cdot m\cdot s}$, and $\Omega_{z'}\sim 10^{10}\,{\rm Hz}$, %or more, 
we can confirm $a_2 a_1 >a_0$. 
Therefore, we can conclude that the declination angle $\theta$ always decays towards zero.

Finally, we check the validity of the assumption of ignoring the non-linear terms in Eq.~(\ref{eq:theta4}).
%whether ignoring non-linear terms in Eq.~(\ref{eq:theta4}) is reasonable. From the estimations above, Eq.~(\ref{eq:theta5}) is roughly given as
Using the same parameter estimate shown above, Eq.~(\ref{eq:theta5}) is approximately rewritten as
%From the estimations above, Eq.~(\ref{eq:theta5}) is roughly given as
\begin{align}
\dddot{\theta}+\frac{B}{I_1}\left(1+\frac{\Omega_{z'}}{\omega_0}\right)\ddot{\theta}+\Omega_{z'}^2\dot{\theta} + \frac{B}{I_1}\Omega_{z'}^2\theta\simeq 0 .
\end{align}
The general solution of this differential equation is written as
%and a general solution of $\theta$ is obtained as
\begin{align}
\theta\simeq C_1 e^{-\frac{B}{I_1}t}+\left(C_2\sin\Omega_{z'}t + C_3\cos\Omega_{z'}t\right) e^{-\frac{B\Omega_{z'}}{2I_1\omega_0}t} ,
\end{align}
where $C_1$, $C_2$, and $C_3$ are determined from the initial conditions 
%of 
for $\theta,\,\dot{\theta},\,\ddot{\theta}$. 
If the initial condition is chosen to satisfy $\dot{\theta}(t=0)\ll \Omega_{z'}\theta(t=0)$ and $\ddot{\theta}\ll \Omega_{z'}^2 \theta(t=0)$, $C_2$ and $C_3$ become much smaller than $C_1$, since $C_1$, $C_2\Omega_{z'}$ and $C_3\Omega_{z'}^2$ are of the same order. 
%If $\dot{\theta}(t=0)\ll \Omega_{z'}\theta(t=0)$ and $\ddot{\theta}\ll \Omega_{z'}^2 \theta(t=0)$, $C_2$ and $C_3$ must be much smaller than $C_1$, since $C_1$, $C_2\Omega_{z'}$ and $C_3\Omega_{z'}^2$ have to be same order. 
Then, the amplitude of $\beta\dot{\theta}/\theta$ and $I_1\ddot{\theta}/\theta$ become much smaller than $a_1\simeq I_1\Omega_{z'}^2$, and neglecting these non-linear terms is justified. 
% 以下の2文は書かなくてもいい気がします　TK
%Then, the amplitude of $\beta\frac{\dot{\theta}}{\theta}$ and $I_1\frac{\ddot{\theta}}{\theta}$ are much smaller than $a_1\simeq I_1\Omega_{z'}^2$, and neglecting these non-linear terms is justified. 
%In this case, non-linear terms $\left(\beta\frac{\dot{\theta}}{\theta},\,I_1\frac{\ddot{\theta}}{\theta}\right)$ becomes important, and Eq.~(\ref{eq:theta5}) becomes improper. 

In summary, we conclude
%can safely argue 
that $\theta$ 
%attenuates 
decays into zero and never shows complex dynamics.
%cannot be larger as time goes on. 
This result indicates that the solution, ${\bm \Omega}=(0,0,\Omega_z)$, which we derived in the main article, is stabilized with respect to  small perturbation tilting the rotation axis.

\bibliography{./reference_supplemental}